\documentclass[11pt,draftcls,peerreview]{IEEEtran}
\usepackage{amssymb}
\usepackage{amsmath,amsthm,amssymb}
\usepackage{graphicx}
\usepackage{color}
\usepackage{algorithm}  
\usepackage{enumerate}
\usepackage[noend]{algorithmic}

\allowdisplaybreaks
\begin{document}
\renewcommand{\IEEEbibitemsep}{0pt plus 2pt}
\makeatletter
\IEEEtriggercmd{\reset@font\normalfont\footnotesize}
\makeatother
\IEEEtriggeratref{1}
\sloppy
\newtheorem{claim}{Claim}
\newtheorem{corollary}{Corollary}
\newtheorem{definition}{Definition}
\newtheorem{example}{Example}
\newtheorem{exercise}{Exercise}
\newtheorem{fact}{Fact}
\newtheorem{lemma}{Lemma}
\newtheorem{note}{Note}
\newtheorem{obs}{Observation}
\newtheorem{problem}{Problem}
\newtheorem{property}{Property}
\newtheorem{proposition}{Proposition}
\newtheorem{question}{Question}
\newtheorem{ru}{Rule}
\newtheorem{solution}{Solution}
\newtheorem{theorem}{Theorem}
\newenvironment{remark}[1]{\textbf{Remark: }}

\def\R{{\rho}}
\newcommand{\A}{{\bf a}}
\newcommand{\bh}{{\bf h}}
\newcommand{\CC}{{\cal C}}
\newcommand{\B}{{\bf b}}
\def\bz{{\bf z}}
\def\bp{{\bf p}}
\newcommand{\I}{{\bf i}}
\newcommand{\p}{{\bf p}}
\newcommand{\q}{{\bf q}}
\newcommand{\x}{{\bf x}}
\newcommand{\y}{{\bf y}}
\newcommand{\z}{{\bf z}}
\newcommand{\M}{{\bf M}}
\newcommand{\N}{{\bf N}}
\newcommand{\X}{{\bf X}}
\def\r {{\bf r}}
\newcommand{\bP}{{\bf P}}
\newcommand{\Q}{{\bf Q}}
\newcommand{\U}{{\bf u}}
\def\d {{\tt d}}
\def\D {{\tt D}}
\def\V {{\tt V}}

\newcommand{\restrict}{{\mathbin{\vert\mkern-0.5mu\grave{}}}}

\newcommand{\glb}{{\tt glb}}
\newcommand{\lub}{{\tt lub}}

\newcommand{\LL}{{\cal L}}
\newcommand{\PP}{{\cal P}}
\newcommand{\pv}{{{\bf p}=(p_1,\ldots , p_n)}}

\newcommand{\qv}{{{\bf q}=(q_1,\ldots , q_n)}}
\newcommand{\Inf}{{\bf t}}
\newcommand{\Sup}{{\bf s}}
\newcommand{\inff}{{\tt t}}
\newcommand{\supp}{{\tt s}}
\def\H {{\cal H}}
\def\l {{\bf l}}

\newcommand{\XX}{\mathcal{X}}
\def\YY{\mathcal{Y}}
\def\FF{\mathcal{F}}
\newcommand{\g}{{\bf g}}
\newcommand{\s}{{sup}}
\def\r {{\bf r}}
\def\t {{\bf t}}
\newcommand{\pcoin}{{coin of bias $p$ }}
\newcommand{\ptree}{{tree of bias $p$ }}
\newcommand{\remove}[1]{}
\newcommand{\commento}[1]{\marginpar{\tiny \flushleft{ #1}}}

\title{Bounds on the Entropy of a Function of  a Random Variable and their Applications}
\author{Ferdinando Cicalese, Luisa Gargano, \IEEEmembership{Member, IEEE} and Ugo Vaccaro, \IEEEmembership{Member, IEEE}%
\thanks{F. Cicalese is with the Dipartimento di Informatica,  Universit\`a di Verona, Verona, 
     Italy (email: ferdinando.cicalese@univr.it), L. Gargano is with the Dipartimento di Informatica,  Universit\`a di Salerno,
		Fisciano (SA), Italy (email: lgargano@unisa.it), 
 and U. Vaccaro is with the Dipartimento di Informatica,  Universit\`a di Salerno,
		Fisciano (SA), Italy (email: uvaccaro@unisa.it).
		This paper is a coalescence and expansion of \cite{CGV,CGV2}.}}

\maketitle
\begin{abstract}
It is well known that the entropy $H(X)$ of a {discrete  random variable $X$}  is always 
greater {than} or equal to the entropy $H(f(X))$ of a function $f$ of $X$, with equality if and only if
 $f$ is one-to-one. In this paper, we give tight bounds on $H(f(X))$
when the function $f$ is not one-to-one, and we illustrate a few  scenarios where 
this matters. As an intermediate step towards our main result, we  
derive  a lower  bound on the entropy of a probability distribution, when only a bound
on the ratio between  the maximal and minimal  probabilities
is  known.
{ The lower bound 
improves on previous results in the literature, and it could  
 find applications outside the present scenario.}
\end{abstract}

\section{The Problem}
Let $\XX=\{x_1,  \ldots ,x_n\}$
be a  finite alphabet, and $X$ be any    random variable (r.v.)
taking values in $\XX$ according to the probability distribution
$\p=(p_1, p_2, \ldots , p_n)$, that is, such that $P\{X=x_i\}=p_i$, 
for $i=1, 2, \ldots , n$. A well known and widely used inequality states
that 
\begin{equation}\label{eq:HX>HfX}
H(f(X))\leq H(X), 
\end{equation}
where $f\colon\XX\to \YY$ is any function defined on $\XX$,  and $H(\cdot)$ denotes  the Shannon
entropy. Moreover, equality holds
in (\ref{eq:HX>HfX}) if and only if the function 
$f$ is one-to-one.
The main   purpose of this paper is to sharpen inequality (\ref{eq:HX>HfX}) by deriving tight bounds on $H(f(X))$ when 
$f$ \emph{is not} one-to-one. More precisely, given the r.v. $X$, an integer $2\leq m<n$, a set $\YY_m=\{y_1,  \ldots ,y_m\}$,
and the family of surjective functions  
$\FF_m=\{f| \;  f\colon\XX\to \YY_m,	\ |f(\XX)|=m \}$,
we want to compute the values
\begin{equation}\label{eq:maxandmin}
\max_{f\in \FF_m}H(f(X)) \qquad \mbox{and} \qquad \min_{f\in \FF_m}H(f(X)).
\end{equation}
While computing the $\min_{f\in \FF_m}H(f(X))$ is  easy, the computation of 
$\max_{f\in \FF_m}H(f(X))$ turns out to be a challenging, but otherwise  consequential, problem. {It is also worth noticing that
maximizing $H(f(X))$ for random functions is trivial, since the value $\log m$ is always achievable}.

\section{The Results}
For any probability distribution $\p=(p_1, p_2, \ldots , p_n)$, with $p_1\geq p_2, \ldots , \geq p_n\geq 0$, and integer 
$2\leq m<n$, let us define the probability distributions $R_m(\p)=(r_1, \ldots ,r_m)$ as follows:
{ if 
$p_1<1/m$ we set $R_m(\p)=(1/m, \ldots ,1/m)$, whereas if $p_1\geq 1/m$
we set  $R_m(\p)=(r_1, \ldots ,r_m)$, where  }
\begin{equation} \label{eq:definition-restriction}
r_i =  \begin{cases}
p_i & \hbox{ for } i = 1, \dots, i^* \cr
\left (\sum_{j=i^*+1}^n p_j\right )/{(m-i^*)} & \hbox{ for } i = i^*+1, \dots, m,
\end{cases}
\end{equation}
and  {$i^*$ is the \emph{maximum} index $i$ such that 
$p_i\geq \frac{\sum_{j=i+1}^n p_j}{m-i}$.}
A somewhat similar operator was introduced in \cite{HY}.

{As suggested by one of the referees, the operator $R_m(\p)$ can also be 
explained in the following way. For a given r.v. $X$ distributed in  the alphabet 
$\XX$ according to  $\p$, the alphabet 
$\XX$ can be partitioned in two sets $\XX_{\mbox{\footnotesize large}}=\{x_1, \ldots , x_{i^*}\}$
and $\XX_{\mbox{\footnotesize small}}=\XX\setminus \XX_{\mbox{\footnotesize large}}$.
Now, a r.v. $\hat{X}$ can be defined to be distributed just like $X$ conditioned on the event that
$X\in \XX_{\mbox{\footnotesize large}}$, and uniformly over $\XX_{\mbox{\footnotesize small}}$,
conditioned on the
event that $X\in \XX_{\mbox{\footnotesize small}}$. When 
the integer $i^*$ is chosen to be the largest possible that ensures that
the probability distribution of $\hat{X}$ remains ordered, then the 
probability distribution of $\hat{X}$ is exactly $R_m(\p)$.
}

\bigskip
We also  define the probability distributions $Q_m(\p)=(q_1, \ldots ,q_m)$ 
in the following way:
\begin{equation} \label{eq:Q}
q_i =  \begin{cases}
\sum_{k=1}^{n-m+1}p_k, \quad  & \hbox{ for } i = 1, \cr
p_{n-m+i}, & \hbox{ for } i=2, \ldots,  m.
\end{cases}
\end{equation}
The following Theorem provides  the results seeked in (\ref{eq:maxandmin}).
\begin{theorem}
For any   r.v.   $X$ 
taking values in the alphabet  $\XX=\{x_1, x_2, \ldots ,x_n\}$ according to the  probability distribution
$\p=(p_1, p_2, \ldots , p_n)$, and for any $2\leq m<n$, it holds that
\begin{equation}\label{max}
\max_{f\in \FF_m}H(f(X))\in \left[H(R_m(\p))-\alpha, H(R_m(\p))\right],
\end{equation}
where
$\alpha=1-({1+\ln(\ln 2)})/{\ln 2}< 0.08608$, and
\begin{equation}\label{min}
\min_{f\in \FF_m}H(f(X))=H(Q_m(\p)).
\end{equation}
\end{theorem}

\bigskip
{Here, with a slight 
abuse of notation, for a probability distribution $\A=(a_1, \ldots , a_t)$ we denote with 
$H(\A)=-\sum_ia_i\log a_i$ the entropy of a discrete r.v. distributed according to 
$\A$. Moreover, with $\log$ we denote the logarithm   in base 2, and with $\ln$ the natural
 logarithm in base $e$}.

\medskip
Therefore, according to Theorem 1, the function $f\in \FF_m$ for which $H(f(X))$ is minimum maps all
the elements $x_1, \ldots , x_{n-m+1}\in \XX$ to a single element, and it is one-to-one
on the remaining elements $x_{n-m+2}, \ldots , x_n$. 

Before proving Theorem 1 and discuss its consequences, we would like to notice that there are quite compelling
reasons why we are unable
to determine  the exact value of the maximum in (\ref{max}), and consequently, the form of 
the function $f\in \FF_m$ that attains the bound. Indeed, 
 computing    the   value $\max_{f\in \FF_m}H(f(X))$ 
is an NP-hard problem. It is easy to understand the difficulty of the problem already
in the simple case $m=2$. To that purpose, consider any function $f\in \FF_2$, that is $f\colon\XX\to\YY_2=\{y_1, y_2\}$,
and let $X$ be any    r.v.
taking values in $\XX$ according to the probability distribution
$\p=(p_1, p_2, \ldots , p_n)$.
Let 
$\sigma_1=\!\!\!\sum_{x\in \XX : f(x)=y_1}P\{X=x\}, \quad \sigma_2=\!\!\!\sum_{x\in \XX : f(x)=y_2}P\{X=x\}.$
Then,  $$H(f(X))=-\sigma_1\log \sigma_1-\sigma_2\log \sigma_2,$$  and it  is maximal 
in correspondence
of a function $f\in \FF_2$ that makes the sums  $\sigma_1$ and $\sigma_2$ 
as much equal as possible. This is equivalent to the well known NP-hard problem  \textsc{Partition} 
on the instance $\{p_1, \ldots , p_n\}$ (see \cite{GJ}).
Actually, we can  prove a  stronger result. 
{We first recall that  problem is said to be strongly NP-hard if it is NP-hard 
even when all of its numerical parameters are bounded by a polynomial in the length of the input \cite{GJ}.
More importantly, any strongly NP-hard optimization problem 
with a polynomially bounded objective function \emph{cannot} have a fully polynomial-time approximation scheme unless 
$P = NP$ \cite{GJ}. 
}

\begin{lemma}\label{strong}
The problem of computing $\max_{f\in \FF_m}H(f(X))$ is strongly NP-hard.
\end{lemma}
\begin{IEEEproof}
The following reduction from the well known {\sc 3-Partition} problem \cite{Weg} proves the result.
We recall that  in the {\sc 3-Partition} problem we are given a set of $N = 3m$ numbers 
$A = \{a_1, \dots, a_N\}$, with $\sum_{i}a_i= m  B$, and such that each number $a_i$ satisfies $\frac{B}{4} < a_i < \frac{B}{2}.$
The question is to decide whether it is possible to partition $A$ into subsets $A_1, \ldots , A_m$,
such that $|A_j|=3$ and $\sum_{a\in A_j}a=B$, for each  $j=1, \ldots m.$
This problem is  known to be strongly NP-complete (see \cite{Weg}, Theorem 7.2.4). 

We will reduce {\sc 3-Partition} to our problem: Let $A$ be an instance of {\sc 3-Partition}, and $N, m$ and $B$ be as above. 
Let $X$ be a random variable taking values in $\{1, \ldots , N\}=[N]$, distributed according to ${\bf p} = (p_1, \dots, p_N)$, where 
$p_i = a_i/(mB)$, for $i=1, \ldots, N$.
 Assume first  that there exists a partition $A_1, \dots, A_m$ of the set  $A$ such that for each $j = 1, \dots, m, $ we have  $|A_j|=3$
and  $\sum_{a \in A_j} a = B$. Let $f$ be  the function $f\colon [N] \mapsto \{1, \ldots , m\}=[m]$ defined by stipulating that 
$f(i) = j$ if and only if $a_i \in A_j$. It is easy to see that $H(f(X)) = \log m.$ 

Conversely, let  $f\colon [N] \mapsto [m]$ be a function  such that $H(f(X)) = \log m.$ This implies that  the random variable
$f(X)$ is equiprobable, i.e., for each $y \in [m]$ we have that 
$$\sum_{i \in [N] \; : \; f(i) = y} p_i = \frac{1}{m}.$$
Let $B_y = \{i \in [N] \, : \, f(i) = y\}.$ Since, by definition, we have $\frac{1}{4m} < p_i < \frac{1}{2m},$ we have that for each $y$ it must hold 
$|B_y| = 3.$ Moreover, we have 
$\sum_{i \in B_{y}} a_i = B m \sum_{i \in B_{y}} p_i  = B.$ Hence, letting $A_y = \{a_i \in A \mid f(i) = y\}$ we have that 
$A_1, \dots, A_m$ is a partition of $A$ into sets of size $3$ with equal total sum. 
Therefore, we can map any instance of {\sc 3-Partition} into an instance of our problem such that there exists a function $f\colon [N] \mapsto [m]$  with 
$H(f(X)) = \log m $ if and only if the {\sc $3$-Partition} instance admits the desired partition.   
\end{IEEEproof}

\bigskip
In this   paper we will also show that  the function $f\in \FF_m$ 
for which $H(f(X))\geq H(R_m(\p))-\alpha$ can be efficiently
determined,  therefore
we also  have  the following important consequence of Theorem 1.
\begin{corollary}\label{cor-approx}
There is a polynomial time algorithm to approximate the NP-hard problem of computing the value
$$\max_{f\in \FF_m}H(f(X)),$$
 with an \emph{additive} approximation factor not greater than  $1-({1+\ln(\ln 2)})/{\ln 2}< 0.08608$.
\end{corollary}
 Under the plausible 
assumption that $P\neq NP$, the strong NP-hardness result proved in Lemma \ref{strong} rules out
 the existence of polynomial time algorithms that, for any values of $\epsilon>0$, 
compute a function $g\in \FF_m$ such that 
$H(g(X))\geq (1-\epsilon)\max_{f\in \FF_m}H(f(X))$.
Therefore, we find it quite interesting  that the problem in question admits  the 
approximation algorithm with the  small additive error mentioned in Corollary 1, since only a
handful  NP-hard optimization problems are known to enjoy this property.
In Section \ref{sub:approx} we will also prove that the polynomial time algorithm 
referred to in Corollary \ref{cor-approx} outputs a solution whose value is at least 
$\frac{10}{11}\Bigl (\max_{f\in \FF_m}H(f(X))\Bigr )$.

\medskip
A key tool  for the proof of Theorem 1 is the following result, proved in 
 Section \ref{proofthm2}.
\begin{theorem}\label{teo-H1}
	Let ${\bf p}=(p_1,p_2, \ldots, p_n)$ be a probability distribution such that  
$p_1\geq  p_2\geq \ldots \geq  p_n>0$. If $ p_1/p_n\leq \R$ then  
\begin{equation}\label{eq:rho}
H({\bf p})\geq \log n - \left(   \frac{\R \ln \R}{\R-1} -1-
                        \ln\frac{\R\ln \R}{\R-1}\right)\frac{1}{\ln 2}.
	\end{equation}											
\end{theorem}

\bigskip
Theorem \ref{teo-H1} improves on   several  papers (see \cite{S+} and 
references therein), that have studied the   problem 
of estimating  $H(\p)$ when only a bound on the ratio $ p_1/p_n$ is 
known. Moreover, besides its application in the proof of the lower bound in (\ref{max}),   
 Theorem \ref{teo-H1} has  consequences of independent interest. 
In particular, in Section \ref{sec:Tunstall} we will show how Theorem \ref{teo-H1}
allows us to provide a new upper bound on the compression rate of Tunstall codes
for discrete memoryless and stationary sources. Our new bound improves the classical result
of Jelinek and Schneider \cite{JS}.

To prove our results, we  use ideas  and techniques  from majorization theory \cite{MO}, 
a mathematical framework that has been  proved to be very  much useful  in 
information theory (e.g., see \cite{CV,CV1,HY,HV} and references therein).

\vspace*{-.41cm}
\section{Some Applications}\label{app1}
Besides its inherent naturalness, the problem of estimating the entropy $H(f(X))$
has several   interesting  consequences. We highlight  some 
of them here. 

\subsection{Clustering}
In the area of clustering \cite{Gan}, one seeks a mapping $f$ (either deterministic or stochastic)
from some data, generated  by a r.v. $X$ taking values in a set $\XX$, 
to ``labels'' in some set $\YY$, where typically 
$|\YY|\ll |\XX|$.  Clusters are  subsets    $C\subseteq \XX$ whose elements are mapped
to a same label $y\in \YY$. 
A widely employed  measure to appraise the goodness of a clustering algorithm is 
the information that the clusters retain towards the original data,
measured by the mutual information $I(X;f(X))$ (see \cite{F+,KMN} and references therein). 
In general, one wants to choose $f$ such that  $|f(\XX)|$ is small
but $I(X;f(X))$ is large. The authors of  \cite{GA} (see also \cite{kur}) proved that, {given} the random
variable $X$, among all mappings  $f$ that maximizes $I(X;f(X))$ (under the
constraint that the cardinality 
$|f(\XX)|$ is fixed) there is a maximizing function $f$ that 
is \emph{deterministic}. 
{This  is essentially a consequence of the fact that the mutual
information is convex in the conditional probabilities}.
Since in the case of deterministic
functions it holds that   $I(X;f(X))=H(f(X))$, it is obvious that 
finding the clustering $f$ of $\XX$ (into a  fixed number $m$ of clusters) 
that maximizes the mutual
information $I(X;f(X))$ is {equivalent} to our problem of finding the
function $f$ that appears in Corollary \ref{cor-approx}.\footnote[3]{In the paper \cite{kur}
the authors consider the    problem of determining the function  $f$ that maximizes
$I(X;f(Y))$, where $X$ is the r.v. at the input of a discrete memoryless channel
 and $Y$ is the corresponding output. Our scenario could be
seen as the particular case when the channel  is noiseless.
However, the results in \cite{kur} do not imply ours  since the authors  give algorithms  only
for  binary input channels (i.e. $n=2$,  that makes the problem completely  trivial in our case). 
Instead, our results are relevant to those of \cite{kur}. For instance, we obtain  that the  general
maximization problem considered in \cite{kur} is strongly NP-hard, a fact unnoticed in \cite{kur}.}
We also remark that  the  problem of 
determining the function $f$ that maximizes the mutual
information $I(X;f(X))$ (under the
constraint that the cardinality 
$|f(\XX)|$ is fixed) has also  been posed in \cite{Dim1,Dim2}.
{Other work that considers the 
general problem of reducing the alphabet size of
a random variable $X$, while trying to preserve the information that it gives 
towards another random variable $Y$, is contained in \cite{NOP,KT}.
Our work seems also related to the well known information bottleneck method \cite{Bot1},
mainly  in the ``agglomerative'' or ``deterministic'' version, see \cite{Bot2,Bot3}.
This connections will be explored elsewhere.}

\subsection{Approximating probability distributions with {low dimensional ones}}
Another scenario where our results directly find applications is the one considered
in \cite{V12}. There, the author considers the problem
of best approximating a probability distribution 
$\p=(p_1, \ldots, p_n)$ with {a lower dimensional}   one
$\q^*=(q^*_1, \ldots , q^*_m)$, $m\leq n$.
The criterion with which one chooses $\q^*$, given $\p$,
is the following. Given arbitrary 
$\p=(p_1, \ldots, p_n)$ 
and $\q=(q_1, \ldots , q_m)$, $m\leq n$, 
define the 
quantity $\D(\p,\q) $ as $2W(\p,\q)-H(\p)-H(\q)$,
where $W(\p,\q)$ is the \emph{minimum} entropy of a bivariate
probability distribution that has $\p$ and $\q$ as marginals.
Equivalently, see \cite[(9)]{V12} $\D(\p,\q) = \min \bigl\{ H(X | Y)  + H(Y | X) \bigr\}$,
where the minimization is with respect to all joint probability distributions of $X$ and $Y$ 
 such that the random variable $X$ is distributed according to  $\p$ and 
the random variable $Y$ according to  $\q$.
{A joint probability distributions of $X$ and $Y$ 
 such that the random variable $X$ is distributed according to \emph{a fixed} $\p$ and 
the random variable $Y$ according to \emph{a fixed} $\q$ is usually called a \emph{coupling} of $X$ and $Y$. 
Couplings (with additional properties) play an important role in information theory questions, e.g., \cite{S}. }

Having so defined the function  $\D$, 
the ``best'' approximation $\q^*$ of $\p$ is chosen as the
probability distributions $\q^*$ with $m$ components that \emph{minimizes} $\D(\p,\q)$,
where the minimization is performed over all probability  distributions $\q=(q_1, \ldots , q_m)$.
The author of \cite{V12} motivates this choice, shows that the function  $\D(\cdot,\cdot)$ 
is a pseudo distance  among probability distributions, and
proves   that $\q^*$ can  be characterized in the following way. 
Given $\p=(p_1, \ldots , p_n)$, 
call  $\q=(q_1, \ldots , q_m)$ an \emph{aggregation} 
of $\p$ into $m<n$ components if there is a partition of $\{1, \ldots , n\}$ into disjoint sets $I_1, \ldots , I_m$
such that $q_k=\sum_{i\in I_k}p_i$, for $k=1, \ldots m$.
In \cite{V12} it is proved that  the vector $\q^*$ that best 
approximate $\p$ (according to $\D$) is the aggregation of $\p$ 
into $m$ components of \emph{maximum entropy}.
 We notice that \emph{any} aggregation $\q$ of $\p$ can 
be   seen as the distribution of the r.v. 
 $f(X)$, where $f$ is  some appropriate non-injective function
 and  $X$ is  a r.v. distributed according to
$\p$ (and, vice versa, any deterministic non-injective $f$ \emph{gives} a r.v. 
$f(X)$ whose distribution is an aggregation of the distribution of the r.v. $X$).
Therefore, 
from Lemma \ref{strong} one gets that  the problem of computing the ``best'' approximation
$\q^*$ of $\p$ is strongly NP-hard.
The author of \cite{V12} proposes   greedy 
algorithms to compute  \emph{sub-optimal} solutions  both to the problem of computing 
 the aggregation of $\p$ 
into $m$ components of {maximum entropy} and to the problem of
computing the probability distributions $\q^*$ with $m$ components that {minimizes} $\D(\p,\q)$.
Notice, however,  that no performance guarantee is given in \cite{V12} for the aforesaid greedy algorithms.
In Section \ref{approx-dist} we will show how 
the bound (\ref{max}) 
allows us to provide an approximation algorithm 
to construct a probability distribution $\overline{\q}=(\overline{q}_1, \ldots , \overline{q}_m)$
such that $\D(\p,\overline{\q})\leq \D(\p,\q^*)+0.08608$, considerably improving 
on 
the result we presented in 
\cite{CGV}, where an approximation algorithm
for the same problem with an additive error of $1$ was provided.

\subsection{Additional relations}
There are  other problems that can be cast in our scenario.
For instance, Baez \emph{et al}. \cite{Baez} give an axiomatic
characterization of the Shannon entropy in
terms of  \emph{information loss}. Stripping
away the Category Theory language of \cite{Baez}, the information loss of a r.v. $X$ amounts
to the difference $H(X)-H(f(X))$, where $f$ is any deterministic function.
Our Theorem 1  allows  to quantify the extreme value of the information
loss of a r.v., when the support of $f(X)$ is known.

{In the paper \cite{CK}  the authors  consider
the problem of constructing the best \emph{summary tree}  of a given weighted tree,
by means of some contractions operations on trees.
{Two type of contractions are allowed: 1) subtrees may be contracted to single node that represent
the corresponding subtrees, 2) multiple sibling subtrees (i.e., subtrees whose roots
are siblings) may be contracted to single nodes representing them. 
Nodes obtained by contracting subtrees have weight equal to the sum
of the node weights in the original contracted subtrees.
} Given a bound on the number of nodes in the 
resulting summary tree, the problem studied in \cite{CK}  is to compute the summary tree of \emph{maximum entropy},
where the entropy of a tree is the Shannon entropy of the normalized node weights. 
This is  a particular case of our problem, when the function $f$ is
not arbitrary but has to satisfy the constraints dictated by the allowed  contractions
operations on trees.

 Another related paper is \cite{D+},
where the authors consider a problem similar to ours, but now $f$ is  
restricted to be a low-degree polynomial and $\p$ is the uniform distribution.}

There is also  a vast literature (see \cite{Ma}, Section 3.3,  and references therein)
 studying  the ``\emph{leakage of a program $P$  [...]  defined as the
(Shannon) entropy of the partition} $\Pi(P)$''  \cite{Ma}. One can easily see that their ``leakage''
is  the same as   the entropy $H(f(X))$, where 
$X$ is the r.v. modeling the program input, and $f$ is the
function describing the input-output relation of the program $P$.
In Section 8 of the same paper the authors study the problem of 
maximizing or minimizing the leakage, in the case the program $P$ is 
stochastic, using standard techniques based on Lagrange multipliers.
They do not consider the (harder) case of deterministic programs (i.e.,
deterministic $f$'s) and our results are likely to be relevant in 
that context.

Our results are also related   
to Rota's entropy-partition theory \cite{rota, rota2}.
Given a ground set $S=\{1, 2, \ldots , n\}$, and a 
partition $\Pi=(C_1, \ldots , C_m)$ into $m$ classes of $S$, the
entropy $H(\Pi)$ of $\Pi$ is defined as
$H(\Pi)=\sum_{i=1}^m \frac{|C_i|}{n}\log \frac{n}{|C_i|}$.
Rota was interested in the decrease (resp. increase) of the entropy 
of a partition under the operation of coarsening (resp., refining) of a partition, where
two or more classes of the partition $\Pi$ are fused into a single class (resp., a  class is split into two or more
new classes). One can see that the decrease of $H(\Pi)$ due to the coarsening operation, for example, can
be quantified by computing the entropy of $H(f(X))$, where $X$ is a r.v. distributed
according to $\left (\frac{|C_1|}{n}, \ldots , \frac{|C_m|}{n}\right)$, and $f$ is an
appropriate function.

Our problem can also be seen as a problem 
of quantizing the alphabet  of a  discrete source  into a smaller one (e.g., \cite{ME}),
and   the goal  is to maximize the mutual information between  
the original source and the quantized one. 
Our results have also relations with those of 	\cite{La+}, where it is considered 
the problems of aggregating data with a minimal information loss.

\section{The Proof of Theorem 1}\label{proofs1}
 We first recall the important concept of
\emph{majorization} among probability distributions.
\begin{definition}\label{defmaj} {\rm \cite{MO}}
Given two probability distributions
$\A=(a_1, \ldots ,a_n)$ and $\B=(b_1, \ldots , b_n)$ with $a_1\geq \ldots \geq a_n\geq 0$ and 
$b_1\geq \ldots \geq b_n\geq 0$, we say that $\A$ is 
{\em majorized} by $\B$, and write  $\A \preceq \B$,
if and only if
$$\sum_{k=1}^i a_k\leq \sum_{k=1}^i b_k, \quad\mbox{\rm for all }\  i=1,\ldots , n.$$
\end{definition}
We will make extensive use of the 
 Schur concavity 
of the entropy function $H$ (see \cite{MO}, p. 101) that says:
\begin{equation}\label{eq:Schur}
 H(\A)\geq H(\B), \quad \mbox{ whenever  } \A\preceq \B. 
 \end{equation}
An important improvement of inequality (\ref{eq:Schur}) was proved in the paper
\cite{HV}, stating that
\begin{equation}\label{eq:Schur2}
 H(\A)\geq H(\B)+D(\B||\A), 
 \end{equation}
{whenever $\A, \B$ have been ordered and $\A\preceq \B$.}
Here  $D(\B||\A)=\sum b_i\log (b_i/a_i)$ is the relative entropy between $\B$ and $\A$.
However, for most of our purposes the inequality (\ref{eq:Schur}) will be sufficient.

\medskip

{One  can also extend the majorization relation to the set of all vectors of finite length
 by padding the shorter vector with zeros and applying  Definition \ref{defmaj}. This is customarily done 
in the literature (e.g., \cite{Nie}). We also notice that this trick does not effect
our results that uses (\ref{eq:Schur}), since adding zeros to a probability distribution $\A$ does
not change the entropy value  $H(\A)$. }

\bigskip
The idea to prove  Theorem 1 is simple. We shall first prove that 
\emph{for any} function $f\in \FF_m$, $m<n$,  and for any r.v. $X$ distributed according
to $\p=(p_1, \ldots , p_n)$,  it holds that $\p$ is majorized by   the probability distribution 
of the random variable $f(X)$. Successively, we will prove that the probability distribution $R_m(\p)$ defined in
(\ref{eq:definition-restriction}) is majorized by  \emph{any} $\q=(q_1, \ldots , q_m)$ 
such that $\p\preceq \q$ (in particular, by the the probability distribution 
of the random variable $f(X)$, with $f\in \FF_m$). These facts, together with the  Schur concavity 
of the entropy function  will prove the upper bound $\max_{f\in \FF_m}H(f(X))\leq H(R_m(\p))$ in (\ref{max}).
We  prove the lower bound in (\ref{max})
by explicitly constructing a function $g\in \FF_m$ such that $H(g(X))\geq H(R_m(\p))-\alpha.$

\medskip
Without loss of generality we assume that \emph{all} the probabilities distributions 
we deal with have been
ordered in non-increasing order. Since we will be working  with functions of probability distributions  that are 
invariant with respect to permutations of the variables, i.e., the Shannon entropy $H(\cdot)$,
this is not a  restriction.  We also  use the majorization 
relationship between vectors of unequal lengths, by properly padding the shorter
one with the appropriate number of $0$'s at the end. The well known  assumption that $0\log\frac{1}{0}=0$
allows us to do that.

Consider   an arbitrary function $f\colon\XX\to\YY$,
$f\in \FF_m$.
Any r.v. $X$ taking values in $\XX=\{x_1, \ldots , x_n\}$,  according to the probability distribution
$\p=(p_1, \ldots, p_n)$,  together with  the function $f$,  naturally induce 
a  r.v. $f(X)$, taking values in $\YY=\{y_1, \ldots , y_m\}$
according to the 
probability distribution 
whose values are given by the expressions
\begin{equation}\label{eq:defy}
\forall y_j\in \YY \qquad P\{f(X)=y_j\}=\sum_{x\in \XX:f(x)=y_j}P\{X=x\}.
\end{equation}
Let $\z=(z_1, \ldots , z_m)$ be the  vector containing the values  
$z_1=P\{f(X)=y_1\}, \ldots , z_m=P\{f(X)=y_m\}$ ordered in non-increasing fashion.
For convenience, we state the following self-evident fact about the
relationships between $\z$ and $\p$.
\begin{claim}
There is a partition of $\{1, \ldots , n\}$ into disjoint sets $I_1, \ldots , I_m$
such that $z_j=\sum_{i\in I_j}p_i$, for $j=1, \ldots m$. 
\end{claim}
\noindent
We will call such a  $\z$  an \emph{aggregation} of $\p$. In the paper \cite{CS} the authors use the
different terminology of \emph{lumping}, our nomenclature is taken from \cite{V12}. 
Given a r.v. $X$ distributed according to $\p$, and \emph{any} function $f\in \FF_m$,  by simply applying 
the definition
of majorization one can see that the (ordered) probability distribution 
of the r.v. $f(X)$ is majorized by  $Q_m(\p)=(q_1, \ldots ,q_m)$,  as defined in 
(\ref{eq:Q}).
Therefore, by invoking the Schur concavity  
of the entropy function $H$ 
we get that 
 $H(f(X))\geq H(Q_m(\p))$. From this, the equality (\ref{min}) immediately  follows.
 
  Denote by 
$$\PP_n=\{\p=(p_1, \ldots , p_n)\in \mathbb{R}^{n}: p_1\geq \ldots p_n\geq 0, \sum_{i} p_i=1\}$$ 
the $(n-1)$-dimensional simplex.
We  need the following  preliminary   results.

{
\begin{lemma}\label{lemma:p<R}
For any $\p\in\PP_n$, $m<n$, it holds that
\begin{equation}\label{eq:p<R}
\p\preceq R_m(\p).
\end{equation}
\end{lemma}
\begin{IEEEproof}
According to Definition \ref{defmaj}, we need to prove that  
\begin{equation}\label{p<Rprova}
\sum_{k=1}^ip_k\leq \sum_{k=1}^ir_k, \mbox{ for } i=1, \ldots , m.
\end{equation}
By the definition (\ref{eq:definition-restriction}) of $R_m(\p)$,  inequalities (\ref{p<Rprova}) are trivially
true for each $i\leq i^*$. { Moreover,  by the  definition of $i^*$ as the largest index $i$ for which
$p_i\geq ({\sum_{j=i+1}^n p_j})/({m-i})$ holds, one has 
 $p_{i^*+1}< ({\sum_{j=i^*+2}^n p_j})/({m-i^*-1})$. Summing up $p_{i^*+1}/(m-i^*-1)$ to both
sides of the previous inequality, one has

$$\frac{\sum_{j=i^*+1}^n p_j}{m-i^*-1} > p_{i^* + 1} + \frac{p_{i^* + 1}}{m-i^*-1} =
p_{i^* + 1} \cdot \frac{m-i^*}{m-i^*-1}.$$

\medskip
Therefore, $p_{i^*+1}<({\sum_{j=i^*+1}^n p_j})/({m-i^*})=r_{i^*+1}$.
Since $\p$
has its  components 
ordered in non increasing  order, 
   one has $p_i\leq r_i= r_{i^*+1}$ also for $i=i^*+1, \ldots , m$.
	
In conclusion, since  we have proved that $p_i\leq r_i$, for all $i=1, \ldots , m$, 
and we also know that $\sum_{k=1}^mr_k=1$, we  get  that  (\ref{p<Rprova}) is proved.}
\end{IEEEproof}

}

\begin{lemma}\label{pprecq}
Let $\p\in \PP_n$, $m<n$,  and  $\q\in \PP_m$ be  \emph{any} aggregation of  $\p$.  Then 
$\p\preceq \q$.
\end{lemma}
\begin{IEEEproof} 
We prove  by induction on $i$ that $\sum_{k=1}^{i}q_k\geq \sum_{k=1}^{i}p_k$.\\
Since ${\bf q}=(q_1, \ldots , q_m)$ is an aggregation of ${\bf p}=(p_1, \ldots , p_n)$, 
one gets that  there is a partition $I_1, \dots, I_m$ of $\{1, \ldots , n\}$
such that $q_i = \sum_{j \in I_i} p_j$ for each $i = 1, \dots, m.$ In particular, 
there exists a subset   $I_j\subseteq \{1, \ldots , n\}$
such that $1\in I_j$. We then  have $q_1\geq q_j\geq p_1$. Suppose now that 
$\sum_{k=1}^{i-1}q_k\geq \sum_{k=1}^{i-1}p_k$. 
If there exist  indices $j\geq i$ and $\ell\leq i$ such that $\ell\in I_j$,
then $q_i\geq q_j\geq p_\ell\geq p_i$,  that implies
$\sum_{k=1}^{i}q_k\geq \sum_{k=1}^{i}p_k$.  
Otherwise,  for each $j\geq i$ and $\ell\leq i$ it holds that $\ell\not \in I_j$. Therefore,
$\{1, \ldots ,i\}\subseteq I_1\cup \ldots \cup I_{i-1}$. This immediately
implies that $\sum_{k=1}^{i-1}q_k\geq \sum_{k=1}^{i}p_k$, from which we obviously
get $\sum_{k=1}^{i}q_k\geq \sum_{k=1}^{i}p_k$.
\end{IEEEproof}

\bigskip
In other words, \emph{for any r.v. $X$ and function} $f$, the probability distribution of $f(X)$
is an aggregation of the probability distribution of $X$.  Therefore, the probability distribution of $f(X)$ 
\emph{always} 
{majorizes}  that  of $X$. As a first consequence, from (\ref{eq:Schur2}) we  get that  
\begin{equation}\label{fX<X}
H(X)\geq H(f(X))+D(\q||\p),
\end{equation}
where $\p$ is the probability distribution of $X$ and $\q$ is the probability distribution of $f(X)$.
This is an improvement of the inequality $H(X)\geq H(f(X))$\footnote{To the best of our knowledge,
the first appearance in print of this inequality in the equivalent form that
 that $H(X)\geq H(Y)$ whenever $H(Y|X)=0$, is in \cite{Sha}. The first paper to 
present the explicit inequality $H(X)\geq H(f(X))$ seems to be \cite{BR}.} that might be of independent interest.
We highlight the inequality (\ref{fX<X}) in the following essentially equivalent formulation.
\begin{corollary}
Let $X$ and $Y$  be \emph{arbitrary}  r.v., distributed according
to $\p$ and $\q$, respectively. Then, if $H(Y|X)=0$ holds, one has  that
$$H(X)\geq H(Y)+D(\q||\p).$$
\end{corollary}

Next Lemma proves that, among all probability distributions 
in $\PP_m$ that majorize a given $\p\in \PP_n$, 
the vector $R_m(\p)\in \PP_m$ defined in (\ref{eq:definition-restriction})
is ``minimal'', according to $\preceq$.
\begin{lemma}\label{Risminimal}
For any $n>m$, $\p\in\PP_n$ and  any $\q\in\PP_m$ it holds that
\begin{equation}\label{eq:Risminimal}
 \p\preceq \q \; \mbox{ implies that  } \; R_m(\p)\preceq \q.
\end{equation}
\end{lemma}
\begin{IEEEproof}
 Consider an \emph{arbitrary} $\q\in \PP_m$ such that $\p\preceq \q$. By definition of $\preceq$ and 
from (\ref{eq:definition-restriction}), we have that $\sum_{k=1}^iq_k\geq \sum_{k=1}^ip_k
=\sum_{k=1}^ir_k$, for $i=1, \ldots , i^*$. It remains to show that 
$\sum_{k=1}^iq_k\geq \sum_{k=1}^ir_k$, also  for $i=i^*+1, \ldots , m$.
Suppose ({\em absurdum hypothesis}), that this is \emph{not} the case and let 
$h\in\{i^*+1, \ldots , m-1\}$ be the smallest integer such that $\sum_{k=1}^hq_k< \sum_{k=1}^hr_k$.
Since $\sum_{k=1}^{h-1}q_k\geq  \sum_{k=1}^{h-1}r_k$,
it follows  that 
$q_h<r_h=(\sum_{j=i^*+1}^n p_j)/({m-i^*})=r$.
Therefore, since  $r_k=r$, for $k\geq i^*+1$,  we have  
$$\sum_{k=h+1}^mq_k=1-\sum_{k=1}^hq_k>1-\sum_{k=1}^hr_k= \sum_{k=h+1}^mr_k=(m-h)r.$$
As a consequence, there exists $j\in \{h+1, \ldots m\}$ such that 
$q_j\geq(\sum_{k=h+1}^mq_k)/({m-h})>r>q_h$, contradicting the fact that 
$q_1\geq q_2\geq \ldots\geq q_m$.
\end{IEEEproof}

\bigskip

 From Lemmas \ref{pprecq} and \ref{Risminimal}, and by applying  the Schur concavity  
of the entropy function $H$, we get the following result.
\begin{corollary}
For any  r.v. $X$ 
taking values in $\XX$ according to a probability distribution
$\p$, and 
for \emph{any} $f\in \FF_m$, 
it holds that
\begin{equation}\label{HF<HR}
H(f(X))\leq H(R_m(\p)).
\end{equation}
\end{corollary}

\bigskip
An equivalent way to say above facts, is that $R_m(\p)$ is the element of $\PP_m$ 
that solves the 
following constrained  maximum entropy problem:
\begin{align}\label{eq-max2}
&\hbox{maximize } H(\q)\\
&\hbox{subject to } \q\in \PP_m  \hbox{ and  } \p\preceq \q. \notag 
\end{align}

Above results imply that 
{
\begin{align}
\max_{f\in \FF_m}H(f(X))\leq H(R_m(\p))&=
                        \sum_{k=1}^{i^*}p_k\log \frac{1}{p_k} 
                           + \left (\sum_{k=i^*+1}^n p_k\right )\log \frac{m-i^*}{\sum_{k=i^*+1}^n p_k},\label{HR}
\end{align}
}

	where $i^*$ has been defined in (\ref{eq:definition-restriction}). 
Therefore, to complete the proof of Theorem 1 we need only  to show that we can construct a  
function $g\in \FF_m$ such that 
\begin{equation}\label{need}
H(g(X))\geq H(R_m(\p))-\left(1-\frac{1+\ln(\ln 2)}{\ln 2}\right),
\end{equation}
or, equivalently, that we can construct    an {aggregation} of $\p$  into $m$ components, 
whose entropy is at least $H(R_m(\p))- \left(1-\frac{1+\ln(\ln 2)}{\ln 2}\right).$
We prove this fact  in  the following lemma.
\begin{lemma}\label{lemma:huffman-prob}
For any $\p=(p_1, \ldots , p_n)$ and $2\leq m<n$, we can construct {an aggregation} ${\bf q}=(q_1, \ldots , q_m)$
of $\p$  
 such that
$$H({\bf q}) \geq H(R_m(\p)) -  \left(1-\frac{1+\ln(\ln 2)}{\ln 2})\right).$$
\end{lemma}
\begin{IEEEproof}
We will assemble the aggregation ${\bf q}$ through the Huffman algorithm.
We first make the following stipulation. To the  purposes of this paper, each  \emph{step} of the 
Huffman algorithm consists in merging the two smallest 
element  $x$ and $y$ of the current probability distribution, 
deleting $x$ and $y$  and substituting them with the single element $x+y$,
and \emph{reordering} the new probability distribution from 
the largest element to the smallest (ties are arbitrarily broken). 
Immediately after the step in which $x$ and $y$ are merged,  \emph{each} element $z$ in the new and reduced
probability distribution that finds itself positioned  at the ``right'' of 
$x+y$ (if there is such a $z$)
 has a value  that satisfies  $(x+y)\leq 2z$ (since, by choice, $x,y\leq z$).
Let ${\bf q} = (q_1, \dots, q_m)$  be the ordered probability 
distribution obtained by executing \emph{exactly} $n-m$ steps 
of the Huffman algorithm, starting from the distribution ${\bf p}$.
Denote by  $i_q$  the maximum index $i$ 
such that for each $j = 1, \dots ,i_q$ the component $q_j$ \emph{has not} been produced by 
a merge operation of the Huffman algorithm. In other word, 
$i_q$ is the maximum index $i$ 
such that for each $j = 1, \dots ,i_q$ it holds that $q_j=p_j$. Notice that we allow $i_q$ to be equal to $0$.
Therefore $q_{i_q+1}$ has been produced by a merge operation. 
At the   step in which the value $q_{i_q+1}$ was created,  it holds that $q_{i_q+1}\leq 2z$,
for any $z$ at the ``right'' of  $q_{i_q+1}$. At later steps, the inequality 
$q_{i_q+1}\leq 2z$ still holds, since elements
at the right of $q_{i_q+1}$ could have  only increased their values.

Let $S = \sum_{k=i_q+1}^mq_k$ be the sum of the last (smallest) $m-i_q$ components of ${\bf q}$. 
The vector  
${\bf q}' = (q_{i_q+1}/S,  \dots ,q_{m}/S)$
is a probability distribution such that the  ratio between its
 largest and its smallest 
component 
is upper bounded by 2.
By Theorem \ref{teo-H1}, with $\rho=2$,  it   follows that 
\begin{equation} \label{equation:DG}
H({\bf q}') \geq \log(m-i_q) - \alpha,
\end{equation}
where $\alpha\leq \left(1-\frac{1+\ln(\ln 2)}{\ln 2}\right)< 0.08608$.
Therefore, we have 
\begin{eqnarray}
H({\bf q}) &=& \sum_{j=1}^{i_q} q_j \log \frac{1}{q_j} + \sum_{j=i_q+1}^{m} q_j \log \frac{1}{q_j}  \label{eq:h-1st}  \label{start}\\
&=& \sum_{j=1}^{i_q} q_j \log \frac{1}{q_j} - S \log S 
 + S \sum_{j=i_q+1}^{m} \frac{q_j}{S} \log \frac{S}{q_j}  \\
&=& \sum_{j=1}^{i_q} q_j \log \frac{1}{q_j} - S \log S + S H({\bf q}')\\
&\geq& \sum_{j=1}^{i_q} q_j \log \frac{1}{q_j} - S \log S  + S (\log(m-i_q) - \alpha) \\
&=& \sum_{j=1}^{i_q} q_j \log \frac{1}{q_j} + \!\!S \log\frac{m-i_q}{S} -  \alpha S  \label{end}\\
&=&  \sum_{j=1}^{i_q} q_j \log \frac{1}{q_j}\! + \!\!\!\sum_{j=i_q+1}^m \frac{S}{m-i_q} \log\frac{m-i_q}{S}  
 - \alpha S  \\
&\geq&  \sum_{j=1}^{i_q} q_j \log \frac{1}{q_j} + \sum_{j=i_q+1}^m \frac{S}{m-i_q} \log\frac{m-i_q}{S}   - \alpha \label{eq:nuova}\\
&=& H\Bigl(q_1, q_2, \dots, q_{i_q}, \frac{S}{m-i_q},  \dots, \frac{S}{m-i_q}\Bigr)  
 -\alpha. \label{eq:h-last} 
\end{eqnarray}

\medskip
{We remark that inequality (\ref{eq:nuova}) holds since $S\leq 1$.}

Let ${\bf q}^* = (q_1, q_2, \dots, q_{i_q}, \frac{S}{m-i_q}, \frac{S}{m-i_q}, \dots, \frac{S}{m-i_q}),$ 
and observe that 
${\bf q}^*$ coincides with ${\bf p}$ in the first $i_q$ components, as it does $\q$.
What we have shown is that 
\begin{equation} \label{h-h*}
H({\bf q}) \geq H({\bf{q}}^*) - \alpha.
\end{equation}

We now observe that $i_q \leq i^*$, where $i^*$ is the index that intervenes in the definition of  our  operator
$R(\p)$ (see (\ref{eq:definition-restriction})). 
In fact, by the definition of $\q$ one has  $q_{i_q} \geq q_{i_q+1} \geq \cdots \geq q_m$, that  also implies  
\begin{equation}\label{eq:hh}
\frac{\sum_{j=i_q+1}^m q_j}{m} \leq q_{i_q+1} \leq q_{i_q} = p_{i_q}.
\end{equation}
Moreover, since the first $i_q$ components of ${\bf q}$ are the same as in ${\bf p}$, we  also have
$\sum_{j = i_{q}+1}^m q_j = \sum_{i_q+1}^n p_j$. This, together with   relation (\ref{eq:hh}), implies
\begin{equation}\label{eq:new}
\frac{\sum_{j=i_q+1}^n p_j}{m} \leq p_{i_q}.
\end{equation}
Equation (\ref{eq:new}) clearly implies $i_q \leq i^*$ since $i^*$ is
by definition,  the maximum index $i$ such that $\sum_{j=i+1}^n p_j \geq (n-i) p_i.$
From the just proved inequality  $i^* \geq i_q$, we have also
\begin{equation} \label{h:inequality}
{\bf q}^* \preceq R_m({\bf p}). 
\end{equation}
Using (\ref{h-h*}), (\ref{h:inequality}),  and the Schur concavity of the entropy function, we get 
$$H({\bf q})\geq H({\bf q}^*)-\alpha\geq H(R_m({\bf p}))-\alpha,$$
thus completing the proof of the Lemma (and  of Theorem 1).
\end{IEEEproof}

{
\subsection{Multiplicative Approximation to $\max_f H(f(X))$}\label{sub:approx}

In this section we  prove the following result.

\begin{theorem} \label{theorem:multiplicative}
For any $\p=(p_1, \ldots , p_n)$ and $2\leq m<n$, we can construct in polynomial time 
{an aggregation} ${\bf q}=(q_1, \ldots , q_m)$
of $\p$ 
 such that
$$H({\bf q}) \geq \frac{10}{11}\max_{f\in \FF_m}H(f(X)).$$
\end{theorem}
\begin{IEEEproof}
Let us consider the algorithm outlined in Lemma \ref{lemma:huffman-prob}, and let ${\bf q}=(q_1, \ldots , q_m)$ be 
the aggregation of $\p$ it produces.
From the series of inequalities (\ref{start})-(\ref{end}), we have that the aggregation $\q$  satisfies the following relations
\begin{align} \label{eq:stronger}
H(\q) &\geq H(\q^*) - \alpha S \nonumber\\
&= H(q_1, q_2, \dots, q_{i_q}, \frac{S}{m - i_q}, \dots, \frac{S}{m - i_q}) - \alpha S, 
\end{align}
where $S = \sum_{j=i_q+1}^m q_j = \sum_{j=i_q+1}^n p_j,$ and, as defined above, $i_q$ is the largest index $k$ such that for each $j \leq k,$
we have $q_j = p_j.$ 

We first observe that if $i_q = m-1$ then $\q$ is optimal, since $i_q \leq i^* \leq m-1$ implies $i_q = i^*$ and $\q = R(\p),$ from which 
$H(\q) = H(R(\p)) \geq \max_{f\in \FF_m} H(f(X)).$

Skipping this trivial case, and the equally trivial case $m = 1,$ we can 
assume that  $$m - i_q \geq 2 \qquad \mbox{and} \qquad m \geq 2.$$
From this, using $0 \leq S \leq 1$  and $\alpha \leq  0.08608$ it follows that 
\begin{equation} \label{ratio-ineq}
\frac{m-i_q}{S} \geq 2 > 2^{11 \alpha}, \qquad \mbox{hence} \qquad \log \frac{m-i_q}{S} > 11 \alpha.
\end{equation}

From the last inequality, we have
\begin{align*}
H(\q^*) &= H(q_1, \dots, q_{i_q}, \frac{S}{m - i_q}, \dots, \frac{S}{m - i_q}) \\
&= \sum_{j=1}^{i_q} q_j \log \frac{1}{q_j}  
+ S \log \frac{m-i_q}{S} \geq S \log \frac{m-i_q}{S} 
> 11 \alpha S.
\end{align*}

Therefore, $$-\alpha S > -\frac{1}{11} H(\q^*),$$
which implies
\begin{align*}
H(\q) &\geq H(\q^*) - \alpha S > \frac{10}{11} H(\q^*) 
      \geq \frac{10}{11} H(R(\p)) \geq \frac{10}{11} \max_{f\in \FF_m} H(f(X)).
\end{align*}
Therefore,  our algorithm produces an aggregation whose entropy is a $\frac{10}{11}$-approximation of the maximum possible.
\end{IEEEproof}

}

\section{The proof of Theorem 2}\label{proofthm2}

We now prove  Theorem \ref{teo-H1}.
Again, we  use  tools from   majorization theory.
Consider an arbitrary  probability distribution  ${\bf p}=(p_1,p_2, \ldots, p_n)$ with  
$p_1\geq  p_2\geq \ldots \geq  p_n>0$ and  $p_1/p_n\leq \R$.
Let us define  the  probability distribution 

\begin{eqnarray}
\bz_\R({\bf p})=(z_1,\ldots,  z_n)&&\\
              =(\underbrace{\R p_n,\ldots, \R p_n}_{i \ \mbox{\scriptsize times}},&&
							                                     \hspace{-0.7truecm} 1-(n+i\R-i-1)p_n,
							                                     \underbrace{p_n, \ldots, p_n}_{n-i-1 \ \mbox{\scriptsize times}}),\nonumber
\end{eqnarray}
where 
 $i=\left\lfloor{(1-np_n)}/{(p_n(\R-1))}\right\rfloor$. 
It is easy to verify that  $p_n\leq 1-(n+i(\R-1)-1)p_n\leq \R p_n$.
\begin{lemma}\label{primo}
Let ${\bf p}=(p_1,p_2, \ldots, p_n)$ with  
$p_1\geq  p_2\geq \ldots \geq  p_n>0$  be any probability distribution  with  $p_1/p_n\leq\R$.
The  probability distribution 
 $\bz_\R({\bf p})$
 satisfies
	${\bf p}\preceq \bz_\R({\bf p}).$
\end{lemma}
\begin{IEEEproof}
For any $j\leq i$, it holds that
$$p_1+\ldots + p_j\leq jp_1\leq j (\R p_n)=z_1+\ldots +z_j.$$
Consider now some  $j\geq i +1$ and  assume by contradiction that 
$p_1+\ldots + p_j >z_1+\ldots+z_j$.
It follows that 
$p_{j+1}+\ldots+p_n<z_{j+1}+\ldots+z_n=(n-j)p_n$. As a consequence 
we get the contradiction  $p_n\leq (p_{j+1}+\ldots+p_n)/(n-j)<p_n$.
\end{IEEEproof}

\medskip
Lemma  \ref{primo} and the Schur concavity of the entropy imply  that $H(\bp)\geq H(\bz_\R({\bf p}))$. 
	We  can therefore prove Theorem 2  by showing    the appropriate  upper  bound on $\log n -H(\bz_\R({\bf p}))$.
\begin{lemma}
	It holds that 
	$$\log n -H(\bz_\R({\bf p}))\leq \left(\frac{\R\ln \R}{\R-1} -1- \ln \frac{\R\ln \R}{\R-1}\right)\frac{1}{\ln 2}.$$
	
\end{lemma}
\begin{IEEEproof}
Consider  the class of  probability distributions of the form

{
$$\bz_\R(x,i)=(\underbrace{\R x,\ldots, \R x}_{i \ \mbox{\scriptsize times}},1-(n+i(\R-1)-1)x,
\underbrace{x, \ldots, x}_{n-i-1\mbox{\scriptsize times}})$$}

having  the first $i$ components equal to  ${\R x}$ and the last $n-i-1$ equal to $x$, for  
suitable  $0\leq x\leq 1/\rho$, and $i\geq 0$ 
such that
 \begin{equation}\label{xin}
1-(n+i(\R-1)-1)x\in[x,\R x).
\end{equation}
Clearly, for $x=p_n$ and $i=\left\lfloor{(1-np_n)}/{p_n(\R-1)}\right\rfloor$ one has    $\bz_\R({\bf p})=\bz_\R(x,i)$, and we can prove the lemma by 
upper bounding the 
maximum (over all  $x$ and $i$) 
of  $\log n -H(\bz_\R(x,i))$.
{ For a fixed $i$, set $A = A(i) = n+i(\R-1)-1$ and let
\begin{align*} \label{fxi}
f(x,i)=& \log n -H(\bz_\R((x,i)) =\log n + i (\R x\log (\R x))\\
 & +  (1-Ax)\log(1-Ax) 
  + (n-i-1)x \log x.
\end{align*}
}
From (\ref{xin}), for any value of  $i\in \{1,\ldots, n-2\}$, one has that
\begin{equation*}
x\in \left(\frac{1}{n+(i+1)(\R-1)},\frac{1}{n+i(\R-1)}\right]
\end{equation*}

{ Let us now study the derivatives of $f(x,i)$ with respect to $x.$ With $A$ defined as above, we have}   
\begin{align*}
f(x,i)=&\log n +i\R x \log(\R x) { +}(1-Ax) \log (1-Ax)\\
        &+(n-i-1)x\log x,\\
\frac{d}{d x}f(x,i)=&i\R\log \R + (i\R -A +n-i-1)\log e \\
                     &+ (i\R +n-i-1)\log x -A\log(1-Ax)\\
										=&i\R\log \R+ A\log x-A\log(1-Ax),\\
\frac{d^2}{d x^2}f(x,i)=& \Bigl(\frac{A}{x}
                          +\frac{A^2}{1-Ax}\Bigr ) \log e.
													\end{align*}
													
Since 
$\frac{d^2}{d x^2}f(x,i)\geq 0$ for 
any value of $x$ in the interval
$ \left(\frac{1}{n+(i+1)(\R-1)},\frac{1}{n+i(\R-1)}\right]$, 
the function is $\cup$-convex in this interval, and it is upper bounded by 
the maximum between the two extrema values $f(1/(n+(i+1)(\R-1)),i)$ and $f(1/(n+i(\R-1)),i)$.

{
We notice now that 
for $x=1/(n+(i+1)(\R-1))$, it holds that
\begin{align*}
\bz_\R(x,i)&=
           (\underbrace{\R x,\ldots, \R x}_{i \ \mbox{\scriptsize times}},1-(n+i(\R-1)-1)x,\underbrace{x, \ldots, x}_{n-i-1 \ \mbox{\scriptsize times}})\\
           &= (\underbrace{\R x,\ldots, \R x}_{i+1 \ \mbox{\scriptsize times}},\underbrace{x, \ldots, x}_{n-1-i \ \mbox{\scriptsize times}}) \\
						&= (\underbrace{\R x,\ldots, \R x}_{i+1 \ \mbox{\scriptsize times}}, 1-(n+(i+1)(\R-1)-1)x, 
						  \underbrace{x, \ldots, x}_{n-i-2 \ \mbox{\scriptsize times}})\\
							&= \bz_\R(x,i+1).
						\end{align*}
and  
$$
f(1/(n+(i+1)(\R-1)),i)=f(1/(n+(i+1)(\R-1)),i+1),$$ 
{  for } $i=1,\ldots, n-2.$
}

Therefore, we can upper bound $f(x,i)$ by  the maximum value among 
\begin{align*}
f(1/(n+i(\R-1)),i)=&\log n+ \frac{i\R}{n+i(\R-1)}\log \R 
                     +\log \frac{1}{n+i(\R-1)},
 \end{align*}
for $i=1,\ldots, n-1$. We now interpret  $i$ as a continuous variable, and 
we differentiate   $\log n+ \frac{i\R}{n+i(\R-1)}\log \R +\log \frac{1}{n+i(\R-1)}$
with respect to $i$.  We get  
\begin{align*}
\frac{d}{d i} & \left(\log n+ \frac{i\R}{n+i(\R-1)}\log \R +\log \frac{1}{n+i(\R-1)}\right)\\
               &=
    \frac{n(\R\log \R-(\R-1)\log e)-i(\R-1)^2\log e}{(n+i(\R-1))^2},
    \end{align*}

\noindent		
that is positive if and only if 
$i\leq \frac{n}{\R-1}\left(\frac{\R\ln \R}{\R-1}-1\right).$
Therefore, the desired upper bound on $f(x,i)$ can be obtained by computing the value of 
$f(\overline{x},\overline{\imath})$, where 
$\overline{\imath}=\frac{n}{\R-1}\left(\frac{\R\ln \R}{\R-1}-1\right)$ and 
$\overline{x}=\frac{1}{n+\overline{\imath}(\R-1)}$. The  value of 
$f(\overline{x},\overline{\imath})$ turns out to be equal to
\begin{align*}
\log n &+
\frac{\frac{n}{\R-1}\left(\frac{\R\ln \R}{\R-1}-1\right)\R\log \R}{n+ n\left(\frac{\R\ln \R}{\R-1}-1\right)}
-\log \left(n+ n\!\left(\frac{\R\ln \R}{\R-1}-1\right)\!\right)\\
&= \frac{\R\log\R(\R\ln\R-\R+1)}{(\R-1)\R\ln \R}-\log\left(\frac{\R\ln\R}{\R-1}\right)\\
&=\frac{\R\ln\R-(\R-1)}{(\R-1)\ln 2} -\log\left(\frac{\R\ln\R}{\R-1}\right) \\
&=\left(\frac{\R\ln \R}{\R-1} -1- \ln\frac{\R\ln \R}{\R-1}\right)\frac{1}{\ln 2}.
\end{align*}
\end{IEEEproof} 

\bigskip
There are several results in the literature that bound from below the entropy
of a probability distribution $\p=(p_1, \ldots, p_n)$, with $p_1\geq  p_2\geq \ldots \geq  p_n>0$,
in terms of the ratio $p_1/p_n$. To the best of our knowledge, the tightest known bound is given in 
\cite{S+}, where it is proven that
if $p_1/p_n\leq 1+2(e^\epsilon-1)+2\sqrt{e^{2\epsilon}-e^\epsilon}$, then
$H(\p)\geq \log n -\epsilon$. We can show  that our bound (\ref{eq:rho}) is  better (see Appendix).

{
\section{An improved  Upper Bound on the Compression Rate of Tunstall Codes}\label{sec:Tunstall}
In variable-to-fixed length encoding of stationary and memoryless discrete sources, the parameter to
minimize is the compression rate $R$ given by
\begin{equation}\label{eq:R}
R=\frac{\lceil \log n\rceil}{E[L]},
\end{equation}
where $n$ is the number of (variable-length) source segments
to be encoded, each with a binary string of length $\lceil \log n\rceil$, and $E[L]$ is the 
average length of the source segments. A classical result by Jelinek and Schneider \cite{JS}
proves that 
\begin{equation}\label{eq:JS}
R\leq \frac{\lceil\log n\rceil}{\log n -\log \rho} \times H(P),
\end{equation}
where $H(P)$ denotes the source entropy and $\rho$ is the reciprocal of the probability of the least probable source symbol.
To prove (\ref{eq:JS}), Jelinek and Schneider \cite{JS} made use of the following two intermediate results:
\begin{equation}\label{eq:JS1}
E[L]=\frac{H(W)}{H(P)},
\end{equation}
where $H(W)$ is the entropy of the leaves of any  parse tree defined by the source
segments, and 
 \begin{equation}\label{eq:JS2}
H(W)\geq \log n -\log \rho,
\end{equation}
where $H(W)$ is the entropy of the leaves of the  parse tree defined by the source
segments produced by the Tunstall algorithm.
Using formul\ae \  (\ref{eq:R}), (\ref{eq:JS1}), the well known fact that in the parse tree produced by the Tunstall algorithm it holds that
the ratio between the largest and smallest probability is upper bounded by the reciprocal of the probability of the least probable source symbol,
  and our Theorem \ref{teo-H1}, we get the following improved upper bound
on the compression rate $R$  
\begin{theorem}\label{th:tunstall}
Consider a stationary and memoryless discrete source whose symbols are compressed by a (variable-to-fixed) Tunstall code 
whose codewords are of length $\lceil \log n \rceil$.
Then, the compression rate of the code satisfies the upper bound 
{
\begin{align}\label{eq:tunstallnew}
R &\leq 
      \frac{ \lceil \log n \rceil H(P) }{ \log n - \left(\frac{\rho (\ln \rho)}{(\rho-1)} - 1 - \ln (\frac{\rho \ln \rho}{\rho-1} ) \right)
      \frac{1}{\ln 2} },
\end{align}
}
where $H(P)$ denotes the source entropy, and $\rho$ is the reciprocal of the probability of the least probable source symbol.
\end{theorem}
}

\begin{figure}[ht]
\includegraphics[width=15truecm,height=7truecm]{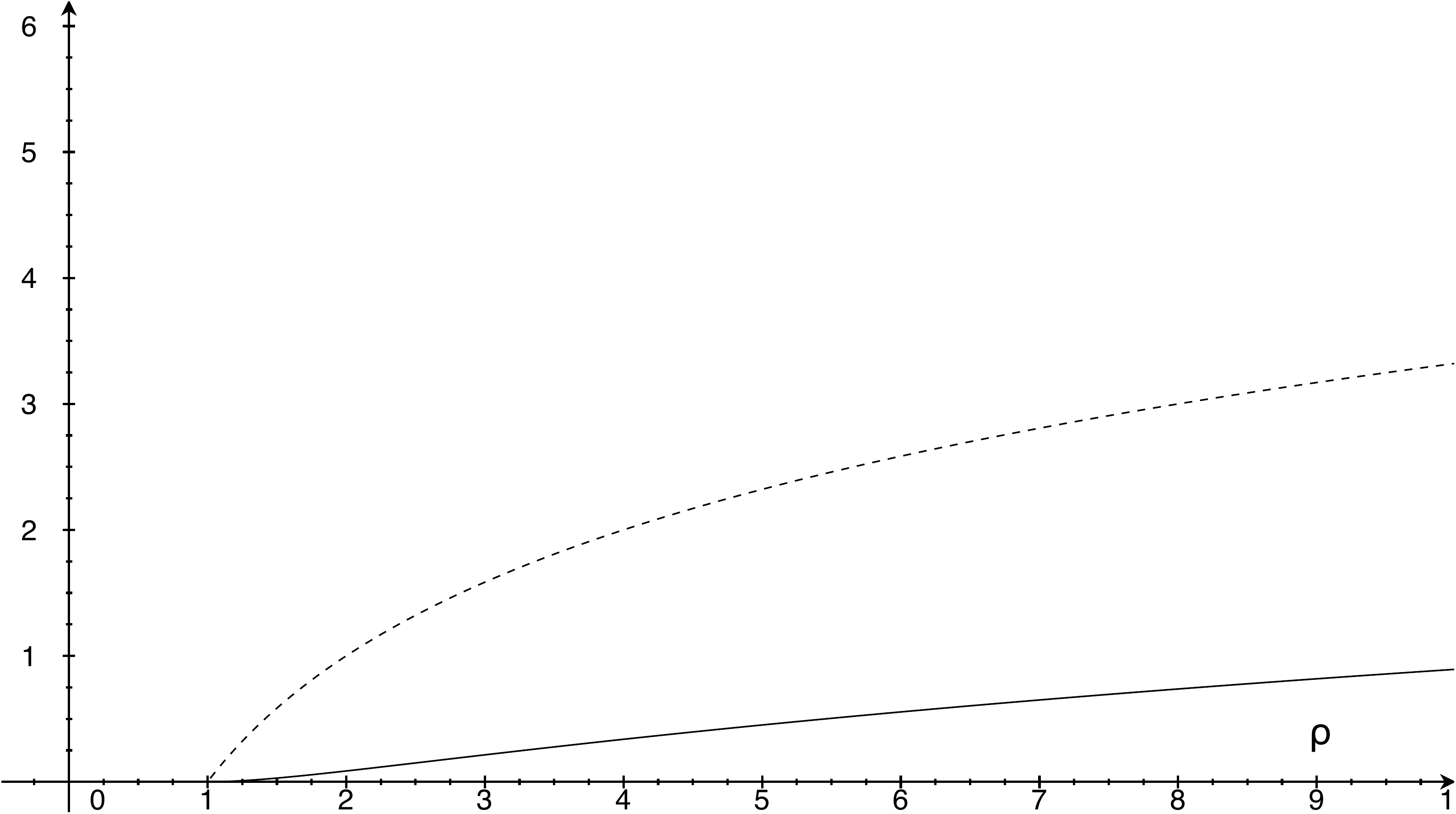}
\caption{Dashed: $\log \R $, solid: $\left(   \frac{\R \ln \R}{\R-1} -1-
                        \ln\frac{\R\ln \R}{\R-1}\right)\frac{1}{\ln 2}$}
												\label{fig1}
\end{figure}

To appreciate how our Theorem 	\ref{th:tunstall} improves on 	(\ref{eq:JS}), we numerically
compare the 			subtractive term in the denominator of 	(\ref{eq:JS}) and of 			(\ref{eq:tunstallnew}).		
In  Figure \ref{fig1} one can see a comparison between the values of 
$\log \R$
and that of 
$\left(   \frac{\R \ln \R}{\R-1} -1-
                        \ln\frac{\R\ln \R}{\R-1}\right)\frac{1}{\ln 2}$. 
The improvement seems significant.

Moreover, our Theorem 2 can also be useful to quantify how close to the uniform probability distribution
is the probability distribution on the leaves of 
a Tunstall tree. This is an important parameter for the 
efficient generations of truly random bits from a biased 
distribution (e.g., \cite{CGV3,MU})
 
\section{Approximating the closest distribution to $\p$, according to the distance \D}\label{approx-dist}
In this section  we  show  how  Theorems 1 and 2 allow us to design an approximation algorithm 
for the second  problem mentioned in Section \ref{app1}, that is, the 
problem of 
constructing  a probability distribution $\overline{\q}=(\overline{q}_1, \ldots , \overline{q}_m)$
such that $\D(\p,\overline{\q})\leq \D(\p,\q^*)+0.08608$, where
$\D(\p,\q)=2W(\p,\q)-H(\p)-H(\q)$,
 $W(\p,\q)$ is the \emph{minimum} entropy of a bivariate
probability distribution that has $\p$ and $\q$ as marginals, and 
$$\q^*=\mbox{arg}\!\!\min_{\q\in \PP_m} \D(\p,\q).$$
 Our algorithm improves
on the result presented in \cite{CGV}, where an approximation algorithm
for the same problem with an additive error of $1$  was provided. 

\smallskip
Let ${\bf q}$ be  the probability distribution
constructed in Lemma \ref{lemma:huffman-prob} and let us recall that the first $i_q$ components of ${\bf q}$ coincide with the first $i_q$ components of ${\bf p}$. 
In addition, for each $i = i_q+1, \dots, m,$ there is a set $I_i \subseteq \{i_q+1, \dots, n\}$ such that
$q_i = \sum_{k \in I_i} p_k$ and the 
$I_i$'s form a partition of $\{i_q+1, \dots, n\},$
(i.e., ${\bf q}$ is an aggregation of $\p$ into $m$ components).
 
We now  build a bivariate probability distribution  ${\M}_q=[m_{ij}]$,
having  $\p$ and ${\bf q}$ as marginals,    as follows:
\begin{itemize}
\item in the first $i_q$ rows and columns, the matrix ${\M}_q$   has non-zero components  only  on the 
diagonal, namely $m_{j\,j} = p_j = q_j$ and $m_{i\, j} = 0$ for any $i, j \leq i_q$ such that $i \neq j$;
\item for each row $i = i_q+1, \dots, m$ the only non-zero elements are the ones in the columns 
corresponding to elements of $I_i$ and precisely, 
for each $j \in I_i$ we set $m_{i \, j} = p_j.$
\end{itemize}

It is not hard to see that $\M_q$ has  ${\bf p}$ and ${\bf q}$ as marginals.
Moreover we have that $H(\M_q) = H({\bf p})$ since by construction the only non-zero components of $\M_q$
coincide with the set of components of ${\bf p}.$ 
Let ${\cal C}(\p, \q)$ be the set of all bivariate probability distribution  
having  $\p$ and ${\bf q}$ as marginals. Recall that
$\alpha=1-({1+\ln(\ln 2)})/{\ln 2}< 0.08608$.
We have that
\begin{eqnarray} \label{eq:V-distance}
\D(\p, \q)&=& \min_{\N \in {\cal C}(\p, \q)} 2H(\N) - H(\p)  - H({\bf q})   \label{eq:V-2} \\
&=& 2 H(\M_q) - H(\p) - H({\bf q}) \label{eq:V-3} \\
&=& H({\bf p}) - H({\bf q}) \label{eq:V-3-1}\\
&\leq& H({\bf p}) - H(R_m(\p)) + \alpha \label{eq:V-5}\\
&\leq& H({\bf p}) - H({\bf q}^*) + \alpha \label{eq:V-5new}\\
&\leq& \D({\bf p}, {\bf q}^*) + \alpha \label{eq:V-6}
\end{eqnarray}
where
(\ref{eq:V-2}) is  the definition of $\D(\p, \q)$;
{(\ref{eq:V-3}) follows from (\ref{eq:V-2})  since  $\M_q \in {\cal C}(\p, \q)$, 
$H(\M_q)= H({\bf p})$, and for all $\N \in {\cal C}(\p, \q)$ one has $H(\N)\geq H(\p)$;}
(\ref{eq:V-3-1}) follows from (\ref{eq:V-3})  because of $H(\M) = H(\p)$; 
(\ref{eq:V-5}) follows from  Lemma \ref{lemma:huffman-prob};
(\ref{eq:V-5new}) follows from  (\ref{eq:V-5}), the known fact 
that ${\bf q}^*$ is an aggregation of $\p$ (see Theorem 8 of \cite{V12}) and  Lemmas 1 and 2.
Finally, the general inequality  $H({\bf a}) - H({\bf b})\leq \D({\bf a}, {\bf b})$
{can be proved in the following  way.
Let $X$ and $Y$ be \emph{arbitrary} random variables distributed according to ${\bf a}$ and ${\bf b}$. Then 
$$H({\bf a}) - H({\bf b})=H(X|Y)-H(Y|X)\leq H(X|Y)+H(Y|X).$$ 
Therefore 
$$H({\bf a}) - H({\bf b}) \leq  \min \bigl\{ H(X | Y)  + H(Y | X) \bigr\}=\D(\p,\q), $$
where the minimization is taken with respect to all joint probability distributions of $X$ and $Y$ 
 such that the random variable $X$ is distributed according to  ${\bf a}$ and 
the random variable $Y$ according to ${\bf b}$. }

\section*{Acknowledgments}
The authors want to thank Associate Editor Professor I. Sason and the anonymous referees 
for  many useful comments,  suggestions  and corrections that went well beyond their call of duty and 
that  helped us to significantly improve and expand  the original submission. In particular, the content
of Section IV.A was inspired by a comment of one of the referees, and the content of Section VI
by suggestions of Professor I. Sason.

\appendix

Let $g(x)=1+2(x-1)+2\sqrt{x^{2}-x}$. The bound in \cite{S+}
tells that
  if   $p_1/p_n\leq g(x)$ for some $x\geq 1$  then 
$H({\bf p})\geq  \log n - \ln x.$
By inverting  the function $y=g(x)$ we get 
$g^{-1}(y)=\frac{(y+1)^2}{4y}$.
Therefore, the bound in \cite{S+}
can be equivalently  stated as follows: If $ p_1/p_n\leq \R$ then 
\begin{equation}\label{simi}
    H({\bf p})\geq  \log n - \ln \frac{(\R+1)^2}{4\R}.
		\end{equation}

The following figure compares   the subtractive term in the bound (\ref{simi})
and the subtractive term in  bound given by our Theorem 2. The interesting regime is for $\rho\geq 1$.
\begin{figure}[ht]
\includegraphics[width=15truecm,height=7truecm]{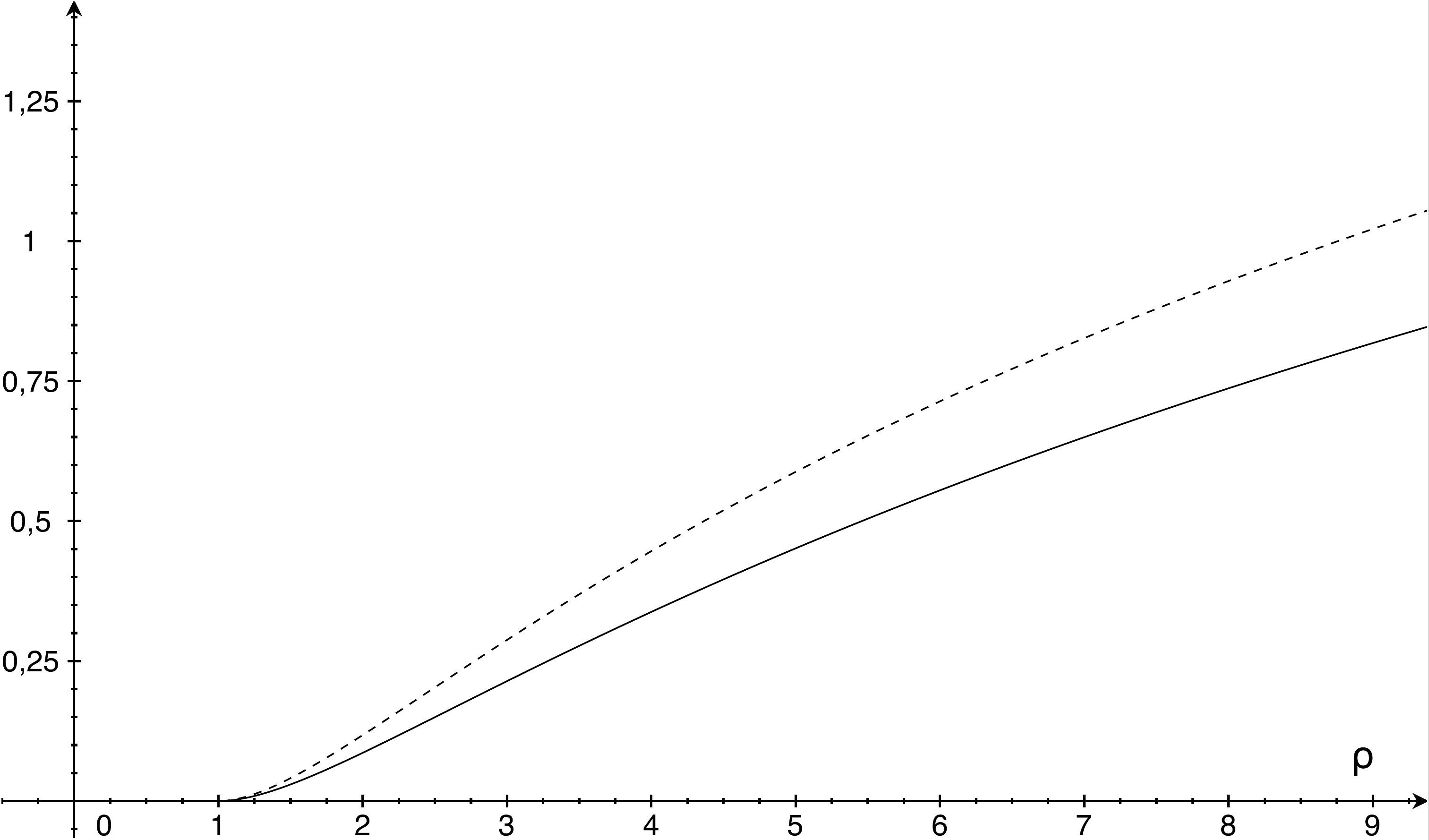}
\caption{Dashed: $ \ln \frac{(\R+1)^2}{4\R}$, solid: $\left(   \frac{\R \ln \R}{\R-1} -1-
                        \ln\frac{\R\ln \R}{\R-1}\right)\frac{1}{\ln 2}.$}
												\label{fig2}
\end{figure}

\begin{thebibliography}{99}
\bibitem{Baez}
 J.C. Baez, T. Fritz and T. Leinster, ``A characterization of entropy in terms of information loss'',
\emph{Entropy}, vol. \textbf{13}, n.11,  1945--1957, 2015.
\bibitem{BR}
J. Balatoni and  A. R\'enyi, ``Remarks on entropy'',  \emph{Publ. Math. Inst. Hung. Acad. Sci.}, vol. \textbf{9},  9–-40, 1956.
	\bibitem{CV}
	F. Cicalese and U. Vaccaro, ``Supermodularity and subadditivity properties of the entropy on the majorization lattice'',
	\emph{IEEE Transactions on	Information Theory},  vol. \textbf{48}, 933--938, 2002.
	\bibitem{CV1}
		F. Cicalese and U. Vaccaro, ``Bounding the average length of optimal source codes via majorization theory'',
		\emph{IEEE Transactions on Information Theory} , vol. \textbf{50}, 633--637, 2004.
	\bibitem{CGV3}  F. Cicalese, L. Gargano, and U. Vaccaro,	``A note on approximation of uniform distributions from variable-to-fixed length codes''. 
\emph{IEEE Transactions on Information Theory}, vol. \textbf{52}, No. 8, pp. 3772--3777, 2006.
\bibitem{CGV4}  F. Cicalese, L. Gargano, and U. Vaccaro,``Information theoretic measures of distances and their econometric applications", 
In: \emph{Proceedings of 2013 International Symposium in Information Theory (ISIT2013)}, pp. 409-413, 2013. 
\bibitem{CGV}  F. Cicalese, L. Gargano, and U. Vaccaro,
``Approximating probability distributions with  short vectors, via  information theoretic distance measures'', 
in: \emph{Proceedings of 2016 IEEE International Symposium on Information Theory (ISIT 2016)}, pp. 1138-1142, 2016. 
\bibitem{CGV2}  F. Cicalese, L. Gargano, and U. Vaccaro,
``$H(X)$ vs. $H(f(X))$'', 
in: \emph{Proceedings of 2017 IEEE International Symposium on Information Theory (ISIT 2017)}, pp. 51--57, Aachen, Germany, 2017.
\bibitem{CK}R. Cole and H. Karloff,``Fast algorithms for constructing maximum entropy summary trees'',
in: \emph{Proc. of 41st International Colloquium  ICALP 2014}, Copenhagen, Denmark, July 8, pp. 332--343, 2014.
\bibitem{CS}
I. Csisz\'ar and P.C. Shields,
`Information Theory and Statistics: A Tutorial'', \emph{Foundations and Trends in Communications and Information Theory}: Vol. \textbf{1}: No. 4, pp. 417-528,
2004.
\bibitem{De} R.C. de Amorim and  C. Hennig, ``Recovering the number of clusters in data sets with noise features 
using feature rescaling factors'',  \emph{Information Sciences}, vol. \textbf{324},
pp.  126–-145, 2015.
\bibitem{Dim1}
A.G. Dimitrov and J.P. Miller,
``Neural coding and decoding: communication channels
and quantization'', \emph{Network: Computation in  Neural Systems}, Vol.  \textbf{12}, pp. 441--472, 2001.
\bibitem{Dim2}
A.G. Dimitrov, J.P. Miller, T. Gedeon, Z. Aldworth, and A.E. Parker,
``Analysis of neural coding through quantization with an information-based distortion measure'',
\emph{Network: Computation In Neural Systems},  Vol. \textbf{14} ,   pp. 151--176, 2003.
\bibitem{D+}Z. Dvir, D. Gutfreund, G. Rothblum, and S. Vadhan, 
``On approximating the entropy of polynomial mappings''. In: \emph{Proceedings of the 2nd Innovations in Computer Science Conference}, pp. 460--475, 2011.
\bibitem{F+}
L. Faivishevsky and J. Goldberger,
``Nonparametric information theoretic clustering algorithm'',
in: \emph{Proceedings of the 27th International Conference on Machine Learning (ICML-10)},
  pp. 351--358, 2010.
	
	\bibitem{Gan}
	G. Gan, C. Ma, and  J. Wu, \emph{Data Clustering: Theory, Algorithms, and Applications}, 
	ASA-SIAM Series on Statistics and Applied
Probability, SIAM, Philadelphia, ASA, Alexandria, VA, 2007.
\bibitem{GJ}  M. R. Garey and D. S. Johnson, \emph{Computers and Intractability: A Guide to the Theory of NP-Completeness},
W. H. Freeman, 1979.
\bibitem{GA}
B.C. Geiger and R.A. Amjad,
``Hard clusters maximize mutual information'',
	\texttt{arXiv:1608.04872 [cs.IT]}
\bibitem{HY} S.W. Ho and R.W. Yeung, ``The interplay between entropy and variational distance'',
\emph{IEEE Trans. Inf..  Theory}, \textbf{56},  pp. 5906--5929, 2010.
\bibitem{HV} S. W. Ho and S. Verd\`u, ``On the interplay between conditional entropy 
and error probability'', \emph{IEEE Trans. Inf.  Theory},  \textbf{56},   pp. 5930--5942, 2010.
\bibitem{JS} F. Jelinek and K.S. Schneider, ``On variable-length-to-block coding'',
\emph{IEEE Transactions on Information Theory}, vol. \textbf{18}, no. 6, pp. 765--774, November 1972.
\bibitem{KT} 	A. Kartowsky and  Ido Tal,
``Greedy-merge degrading has optimal power-law'', 
in: \emph{Proceedings of 2017 IEEE International Symposium on Information Theory (ISIT 2017)}, pp. 1618--1622, Aachen, Germany, 2017. 

\bibitem{KMN}
M. Kearns,  Y. Mansour, and A. Y. Ng, ``An information-theoretic analysis of hard and soft assignment methods for clustering.'' 
In: \emph{Learning in graphical models}. Springer Netherlands, pp. 495--520, 1998.
	\bibitem{KSS15}
	M. Kova\v{c}evi\'c, I. Stanojevi\'c, and V. Senk,  ``On the entropy of couplings'',  \emph{Information
	and Computation}, vol. \textbf{242}, pp. 369--382,  2015.
\bibitem{kur} B.M. Kurkoski,and H. Yagi, ``Quantization of binary-input discrete
memoryless channels'', \emph{IEEE Transactions   Information Theory},  vol. \textbf{60},  4544 -- 4552,  2014.
\bibitem{NOP} B. Nazer,   O. Ordentlich, and   Y. Polyanskiy,
``Information-distilling quantizers'', in: \emph{Proceedings of 2017 IEEE International Symposium on Information Theory (ISIT 2017)}, pp. 96--100, Aachen, Germany, 2017. 
\bibitem{rota} G.-C. Rota, ``Twelve problems in probability no one likes to bring up'', in:  \emph{Algebraic Combinatorics and Computer Science}, 
H. Crapo \emph{et al}. (eds.), pp. 57--93, Springer-Verlag, 2001.
 \bibitem{rota2}J.P.S. Kung, G.-C. Rota, and  C.H. Yan, \emph{Combinatorics: The Rota Way}, Cambridge University Press, 2009.
\bibitem{La+} R. Lamarche-Perrin, Y. Demazeau, J.-M. Vincent, ``The best-partitions problem:
How to build meaningful aggregations'', \emph{2013 IEEE/WIC/ACM Inter. Conf. on Web Intell.  and Intelligent Agent Techhology}, 309--404, 2013.
\bibitem{Ma}
P. Malacaria and J. Heusser,
``Information theory and security: Quantitative information flow'', in:
 Aldini A., Bernardo M., Di Pierro A., Wiklicky H. 
(eds) \emph{Formal Methods for Quantitative Aspects of Programming Languages. SFM 2010}. Lecture Notes in Computer Science, vol. \textbf{6154}. Springer, Berlin, Heidelberg, 2010.
\bibitem{MO} A.W. Marshall, I. Olkin,   B. C. Arnold, \textit{Inequalities: Theory of
        Majorization and Its Applications}, 
         Springer, New York, 2009.
				\bibitem{MU}
				 M. Mitzenmacher  and  E. Upfal,
				\emph{Probability and Computing: Randomized Algorithms and Probabilistic Analysis},
				Cambridge University Press, 2005.
				\bibitem{ME} D. Muresan and  M. Effros, ``Quantization as histogram segmentation: Optimal scalar quantizer design in network systems'',
				\emph{IEEE Transactions on Information Theory},  vol. \textbf{54}, 344--366, 2008.
				\bibitem{Nie} M.A. Nielsen and G. Vidal, ``Majorization and the interconversion of bipartite states'',
				\emph{Quantum Information and Computation}, vol. \textbf{1}, no. 1, 76--93, 2001.
	\bibitem{S} I. Sason, ``Entropy bounds for discrete random variables via maximal coupling,'' \emph{IEEE Trans.
on Information Theory}, Vol. \textbf{59}, no. 11,  7118--7131, 2013.

\bibitem{Sha}
C. Shannon, ``The lattice theory of information'',  \emph{Transactions of the IRE Professional Group on Information Theory},
Vol.  \textbf{1},  pp. 105--107, 1953.
\bibitem{S+} S. Simic,  ``Jensen's inequality and new entropy bounds.'' 
\emph{Appl. Math. Letters},
 \textbf{22},  1262--1265,   2009.
\bibitem{Bot2} N. Slonim and N. Tishby, ``Agglomerative information bottleneck'', in: \emph{Advances in Neural Information Processing Systems 12 (NIPS 1999)}, pp. 617--623.
Denver, CO, USA, 1999.
\bibitem{Bot3} D.J. Strouse and D.J. Schwab, ``The Deterministic information bottleneck'', \emph{Neural Computation}, vol. \textbf{29}, no. 6, pp. 1611--1630,
 2017.
\bibitem{Bot1}
N. Tishby, F.  Pereira, and W. Bialek,  ``The Information bottleneck method'',
in:  \emph{The 37th annual Allerton Conference on Communication, Control, and Computing}, pp. 368–-377, Monticello, IL., USA, 1999.
\bibitem{V12}
	M. Vidyasagar, 
	``A metric between probability distributions on finite sets of different cardinalities and applications to order reduction'', 
\emph{IEEE Transactions on	Automatic Control},    vol. \textbf{57},  2464--2477, 2012.

\bibitem{Weg} I. Wegener, \emph{Complexity Theory}, Springer-Verlag, 2005.
\end{thebibliography}
\end{document}